\documentclass[runningheads]{llncs}
\usepackage{graphicx}
\usepackage{url}  
\usepackage[it]{subfigure}
\usepackage{array}
\newcolumntype{L}[1]{>{\raggedright\let\newline\\\arraybackslash\hspace{0pt}}m{#1}}
\newcolumntype{C}[1]{>{\centering\let\newline\\\arraybackslash\hspace{0pt}}m{#1}}
\newcolumntype{R}[1]{>{\raggedleft\let\newline\\\arraybackslash\hspace{0pt}}m{#1}}
\begin{document}
\title{Quantifying Polarization on Twitter: the Kavanaugh Nomination}
\titlerunning{Quantifying Polarization on Twitter}
%
\author{Kareem Darwish 
}
\authorrunning{K. Darwish}
\institute{Qatar Computing Research Institute, HBKU, Doha, Qatar \\ 
\email{kdarwish@hbku.edu.qa}\\
}
\maketitle              
\begin{abstract}
This paper addresses polarization quantification, particularly as it pertains to the nomination of Brett Kavanaugh to the US Supreme Court and his subsequent confirmation with the narrowest margin since 1881. Republican (GOP) and Democratic (DNC) senators voted overwhelmingly along party lines.  In this paper, we examine political polarization concerning the nomination among Twitter users.  To do so, we accurately identify the stance of more than 128 thousand Twitter users towards Kavanaugh's nomination using both semi-supervised and supervised classification. Next, we quantify the polarization between the different groups in terms of who they retweet and which hashtags they use.  We modify existing polarization quantification measures to make them more efficient and more effective. We also characterize the polarization between users who supported and opposed the nomination.

\end{abstract}
\section{Introduction}
On October 6, 2018, the US senate confirmed Brett Kavanaugh (BK) to become a justice on the US Supreme Court with a 50 to 48 vote that was mostly along party lines.  This was the closest successful confirmation to the court since the Stanley Matthews confirmation in 1881\footnote{\url{https://www.senate.gov/pagelayout/reference/nominations/Nominations.htm}}.  Political polarization was clearly evident in the US Senate between Republicans, who overwhelmingly voted for Kavanaugh, and Democrats, who overwhelmingly voted against him.  In this paper, we wanted to quantify the political polarization between Twitter users, who voiced their opinion about the nomination. Quantification involved: a) collecting topically relevant tweets; b) ascertaining the stances of users; and c) properly quantifying polarization. For data collection, we collected more than 23 million tweets related to BK's nomination, and we semi-automatically tagged more than 128 thousand Twitter users as supporting or opposing his confirmation. We initially manually tagged a small set of active users, then performed label propagation based on which tweets they retweet, and lastly used supervised classification to tag a greater number of users based on the accounts they retweeted.  As for quantification, we modified two existing polarization quantification measures, namely Random Walk Controversy (RWC) and  Embedding Controversy (EC) measures that were shown to be indicative of polarization \cite{garimella2018quantifying}. Given a graph of connected users, where users are the nodes and the weights of the edges are the similarities between users, RWC is computed based on the likelihood that a shorter graph traversal can be made from a random user to a prominent user with the same stance or to a prominent user with a different stance.  EC maps users into a lower dimensional space and then computes a ratio of distances between users with similar stances and users with different stances.  
Due to the high computational complexity of the measures, we use user samples, and we estimate the stability of the measures across: multiple samples and different sample sizes.  Further, we modify the original measures reported in the literature to make them more robust. 
We apply the modified measures on the Twitter users who actively discussed BK's nomination.  We show strong polarization between both camps particularly in terms of the accounts that users retweet. 

Next, we highlight polarization by bucketing hashtags, retweeted accounts, and cited websites according to how strongly they are associated with those who supported or opposed the nomination.  We show that the polarization of Twitter users caused them to retweet different accounts, cite different media sources, and use different hashtags. 
In doing so, we highlight some of the main differences between both groups. The contributions of the paper are:
\begin{itemize}
    \item We showcase effective semi-supervised user labeling that combines both multiple label propagation iterations and supervised classification. We show the effectiveness of this combination in tagging more than 128 thousand users using a very small initial set of manually tagged users. 
    \item We experimented with two reportedly effective polarization quantification measures, namely EC and RWC, elucidate their shortcomings on our dataset, and propose modifications to make them more robust.
    \item We analyze users who were vocal on Twitter concerning the BK nomination.  We characterize them in terms of the hashtags they use, the accounts they retweet, and the media sources that they cite.
\end{itemize}
\vspace{-3pt}
\section{Background}
\subsection{Stance Detection}
Given the ubiquity of social media use, stance detection, which involves identifying the position of a user towards an entity or a person, is emerging as a problem of increasing interest in the literature. We are specifically interested in stance detection on Twitter. Stance detection can be performed at user-level or at statement-level.  For either case, classification can be performed using a variety of features such as textual features (e.g. words or hashtags), interaction-level features, (e.g. relationships and retweeted accounts), and profile-level features (e.g. user location and name) \cite{borge2015content,magdy2016isisisnotislam,magdy2016failedrevolutions,weber2013secular}. Typically, interaction-level features yield better results \cite{magdy2016isisisnotislam}.  In a supervised setting, an initial set of statements and/or users are tagged with their stance, which is used to train a classifier \cite{borge2015content,magdy2016isisisnotislam}. This is appropriate for user-level and statement-level classification \cite{mohtarami2018automatic}.  Alternatively, so-called label propagation is used to propagate labels in a network based on interactions between users such as follow or retweet relationships \cite{borge2015content} or the retweeting of identical tweets \cite{kutlu2018devam,magdy2016isisisnotislam}.  Label propagation has been shown to produce highly accurate results. In this work, we manually tag an initial set of users, employ label propagation, and then use the output labels from label propagation to perform supervised classification. More recent work has focused on unsupervised stance detection that involves creating a user-similarity network based on interaction-level features, and then combines dimensionality reduction, such as Uniform Manifold Approximation and Projection (UMAP) or force-directed (FD) graph placement, with clustering, such as mean shift, to identify users who are strongly associated with specific stances \cite{darwish2019unsupervisedStance}.  
\vspace{-3pt}
\subsection{Quantifying Polarization}
\label{sec:quantifyingPolarization}
Quantifying polarization can help ascertain the degree to which users are separable based on their stances and how far apart they are. Though multiple measures have been suggested for quantifying polarization, research on establishing widely accepted effective measures is still work in progress. Guerra et al. \cite{guerra2013measure} introduced a polarization measure that relies on identifying popular nodes that lie at the boundary between different communities, where strong polarization is indicated by the absence of such nodes.  Morales et al. \cite{morales2015measuring} proposed a metric that measures the relative sizes of groups with opposing views and the distance between their ``centers of gravity''.  Garimella et al. explored a variety of controversy quantification measures to ascertain their efficacy \cite{garimella2018quantifying}. The measures rely on random graph walks, network betweenness, and distances in embedding spaces.  Given their reported success, we employ so-called Random Walk Controversy (RWC) and Embeddings Controversy (EC) measures \cite{garimella2018quantifying} in this paper.  Given the most connected nodes in a graph, RWC uses the maximum likelihood estimates that a random element in one class would reach one of the most connected nodes in its class first or one of the most connected nodes in the other class first. EC maps users into a lower dimensional space and then computes a ratio of the inter- and intra-class distances between users. We propose modifications to both measures to make them more computationally efficient and more robust. As we show in the paper, aside from direct measures of polarization, the effects of polarization can be observed in data \cite{borge2015content,conover2011political,morales2015measuring,weber2013secular}.  For example, projecting users who engage in discussing a polarized topic on to a lower dimensional space can help visualize such polarization and improve subsequent clustering \cite{darwish2019unsupervisedStance,garimella2018quantifying}.  Further, polarized groups tend to share content from different media sources and influencers on social media, and often use different words and hashtags.    
\vspace{-3pt}
\subsection{Topic Timeline}
On July 9, 2018, Brett Kavanaugh (BK), a US federal judge, was nominated by the US president Donald Trump to serve as a justice on the US supreme court to replace outgoing Justice Anthony Kennedy\footnote{\url{https://en.wikipedia.org/wiki/Brett_Kavanaugh}}. His nomination was marred by controversy with Democrats complaining that the White House withheld documents pertaining to BK's record and later a few women including a University of California professor accused him of sexual assault\footnote{\url{https://www.nytimes.com/2018/09/26/us/politics/brett-kavanaugh-accusers-women.html}}. The accusations of sexual misconduct led to a public congressional hearing on September 27, 2018 and a subsequent investigation by the Federal Bureau of Investigation (FBI).  The US Senate voted to confirm BK to a seat on the Supreme Court on October 6 with a 50--48 vote, which mostly aligned with party loyalties.  BK was sworn in later the same day.

\section{Data Collection}
We collected tweets pertaining to the nomination of BK in two different time epochs, namely September 28-30, which were the three days following the congressional hearing concerning the sexual assault allegation against BK, and October 6-9, which included the day the Senate voted to confirm BK and the subsequent three days.  We collected tweets using the twarc toolkit\footnote{\url{https://github.com/edsu/twarc}}, where we used both the search and filtering interfaces to find tweets related to the nomination. The keywords we used included BK's name (\textit{Kavanaugh}), his main accuser (\textit{Ford}), the names of the members of the Senate's Judiciary Committee (\textit{Blasey}, \textit{Grassley}, \textit{Hatch}, \textit{Graham}, \textit{Cornyn}, \textit{Lee}, \textit{Cruz}, \textit{Sasse}, \textit{Flake}, \textit{Crapo}, \textit{Tillis}, \textit{Kennedy}, \textit{Feinstein}, \textit{Leahy}, \textit{Durbin}, \textit{Klobuchar}, \textit{Coons}, \textit{Blumenthal}, \textit{Hirono}, \textit{Booker}, and \textit{Harris}), and the words \textit{Supreme}, \textit{judiciary}, \textit{Whitehouse}.  Though some of these terms are slightly more general (e.g. Ford or Whitehouse), potentially leading to non-relevant tweets, the public focus on the nomination during the collection period would have minimized such an effect.  The per day breakdown of the collected tweets is as follows:
\begin{center}
\small{
\begin{tabular}{C{1.4cm}|C{1.4cm}|C{1.4cm}|C{1.4cm}|C{1.4cm}|C{1.4cm}|C{1.4cm}|C{1.5cm}}
 28-Sep & 29-Sep & 30-Sep & 6-Oct & 7-Oct & 8-Oct & 9-Oct & Total \\ \hline
5,961,549 & 4,815,160 & 1,590,522 & 2,952,581 & 3,448,315 & 2,761,036 & 1,687,433 & 23,216,596 \\
\end{tabular}
}
\end{center}

In all, we collected 23 million tweets that were authored by 687,194 users.  Our first step was to accurately label as many users as possible by their stance as supporting (SUPP) or opposing (OPP) BK's confirmation.  The labeling process was done in three steps, namely:\\
1. \textbf{Manual labeling of users.} We manually labeled 43 users who had the most number of tweets in our collection.  The labeling was performed by one annotator who is well-versed in American politics. Of them, the SUPP users were 29 compared to 12 OPP users.  As for the two remaining users, one was neutral and the other was a spammers.  \\
2. \textbf{Label propagation.} Label propagation automatically labels users based on their retweet behavior \cite{darwish2017predicting,kutlu2018devam,magdy2016isisisnotislam}.  The intuition behind this method is that users that retweet the same tweets on a topic most likely share the same stance. Given that many of the tweets in our collection were actually retweets or duplicates of other tweets, we labeled users who retweeted 15 or more tweets that were authored or retweeted by the SUPP group or 7 or more times by OPP group and no retweets from the other side as SUPP or OPP respectively.  We elected to increase the minimum number for the SUPP group as they were over represented in the initial manually labeled set. Such manual tweaking is one of the drawbacks of label propagation \cite{darwish2019unsupervisedStance}. We iteratively performed such label propagation 4 times, which is when label propagation stopped labeling new accounts.   After the last iteration, we were able to label 65,917 users of which 26,812 were SUPP and 39,105 were OPP. Since we don't have golden labels to compare against, we opted to spot check the results. Thus, we randomly selected 10 automatically labeled accounts, and all of them were labeled correctly. We do more thorough checks later.  This labeling methodology naturally favors users who are more opinionated and vocal about a topic and hence hold strong views. \\ 
3. \textbf{Retweet-based classification.} We used the labeled users to train a classification model to guess the stances of users who retweeted at least 20 different accounts, which were users who were actively tweeting about the topic.  For classification, we used the FastText classification toolkit, which is an efficient deep neural network classifier that has been shown to be effective for text classification \cite{joulin2016bag}.  We used the Twitter handles of the accounts that each user retweeted as features.  Strictly using the retweeted accounts has been shown to be effective for stance classification \cite{magdy2016isisisnotislam}.  To keep precision high, we only trusted the classification of users where the classifier was more than 90\% confident.  In doing so, we increased the number of labeled users to 128,096, where 57,118 belonged to the SUPP group 
and 70,978 belonged to the OPP group. 
We manually and independently labeled 100 random users, 50 from each class, who were automatically tagged, and manual and automatic labeling agreed for 96 of them. It is noteworthy that the relative number of SUPP to OPP users in not necessarily representative of real life.  

\section{Quantifying Polarization}
Given the labeled users, we attempted to quantify the polarization between users given the aforementioned EC and RWC measures. Both measures range between 0 (no polarization) and 1 (extreme polarization).  Due to the computational complexity of both measure, we resorted to computing the measures on random samples of users.  We wanted to ascertain: a) the sensitivity of the measures to the size of the samples; and b) to the stability of the measure across different samples. \\
Given a graph of users, as nodes, and edges between them, weighted by the similarity between users, RWC is based on the maximum likelihood estimates that a random element in one class would reach via a graph random walk one of the most connected nodes in its class first or one of the most connected nodes in the other class. The formulation of the score is: $RWC = P_{AA} P_{BB} - P_{AB} P_{BA}$, where $A$ and $B$ are different classes and $P_{XY}$ is the probability that a random node in $X$ would reach a highly connected node in $Y$. We selected the top most connected users in each class and an equal number of random users from each.  To compute cosine similarity, each user was represented by a vector of all the hashtags that they have used (H) or all the accounts that they have retweeted (R).  We modified the method of computing RWC in one important way, compared to what is described in \cite{garimella2018quantifying}. Namely, instead of relying on the minimum number of hops, traversed edges, required to link two users, for which we would have needed to ascertain a minimum threshold for a link, we opted to use the minimum product of the weights of the edges to be traversed to link two users.  
This is akin to computing the shortest path in a graph, and relieves us from trying to determine appropriate thresholds and, as we show later, leads to more consistent results.  We computed the score of a full path as the product of cosine similarities along all the edges between the random node and one of the highly connected nodes.

EC on the other hand relies on projecting nodes based on their similarity into a lower dimensional space, and then computing the average distances between members of the same class (inter-class) or members of different classes (intra-class). EC is computed as: $EC = 1 - \frac{d_A + d_B}{2 d_{AB}}$, where $d_A$ and $d_B$ are the average inter-class distances and $d_{AB}$ is the average intra-class distance.  
In the work of \cite{garimella2018quantifying}, they used force directed graph (FD) placement to perform dimensionality reduction.  In this work, we use both FD as well as UMAP, which is more aggressive than FD in projecting similar users closer together while pushing dissimilar users further apart.  
Once users are projected and hopefully separated in the lower dimensional space, we used Euclidean distance between them to measure average inter- and intra-class distances.  Due to the large number of users, it was computationally prohibitive to 
project all users based on their similarity.  To overcome the computational issue, we opted to use a sample of users to compute EC.  However, we wanted to ascertain the sensitivity of EC to sample size and its sensitivity to random user selection.  Thus, for every sample size, we sampled users 5 times and we computed the average EC and standard deviation across all samples.  Table \ref{tab:QuantifyingPolarizationParameters} lists the parameters we used for both RWC and EC.  For sample sizes per class, we experimented with 500, 1,000, 2,610, and 5,000.  Given the size of our set of labeled users, roughly 128k, 2,610 is the size of a representative random sample of users with 99\% confidence and a margin of error $= \pm 2.5$.\footnote{Calculated using \url{https://surveysystem.com/sscalc.htm}}  We compared two sets of labeled users, namely after label propagation alone and after both label propagation and supervised classification.

\begin{table}[ht]
    \centering
    \caption{Parameters used to for polarization quantification measures.}
    \begin{tabular}{l|l}
        \multicolumn{2}{c}{RWC} \\ \hline
         Parameter & Values \\ \hline
         Top connected accounts per class & 20 \\ 
         Sample size per class & 500, 1,000, 2,650, 5,000 \\
         Similarity feature & Hashtags (H), Retweets (R) \\ \hline
         \multicolumn{2}{c}{EC} \\ \hline
         Parameter & Values \\ \hline
         Sample size per class & 500, 1,000, 2,610, 5,000 \\
         Dimentionality reduction method & UMAP, FD \\
         Similarity feature & Hashtags (H), Retweets (R) \\ \hline
    \end{tabular}
    \label{tab:QuantifyingPolarizationParameters}
\end{table}

\begin{figure*}[h]
    \centering
    \includegraphics[width=0.6\linewidth]{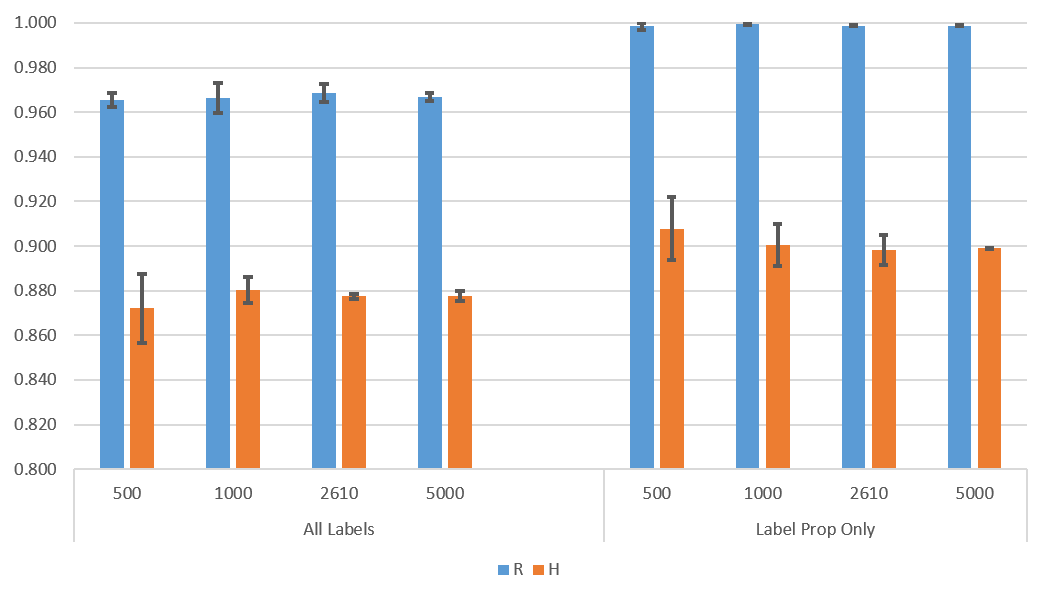}
    \caption{Comparing different setups for RWC with error bars representing standard deviation -- y-axis is the average RWC across 5 different samples.}
    \label{fig:resRWC}
\end{figure*}

\begin{figure*}[h]
    \centering
    \includegraphics[width=.8\linewidth]{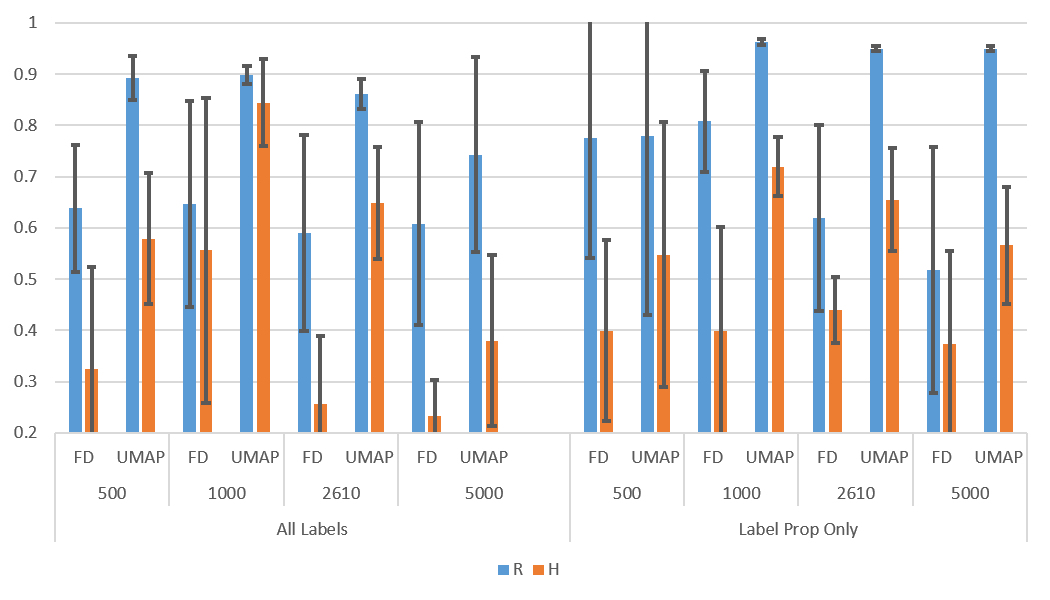}
    \caption{Comparing different setups for EC with error bars representing standard deviation -- y-axis is the average EC across 5 different samples.}
    \label{fig:resEC}
\end{figure*}

\begin{figure}[h]
    \centering
         \subfigure[FD using R]{\includegraphics[width=0.43\linewidth]{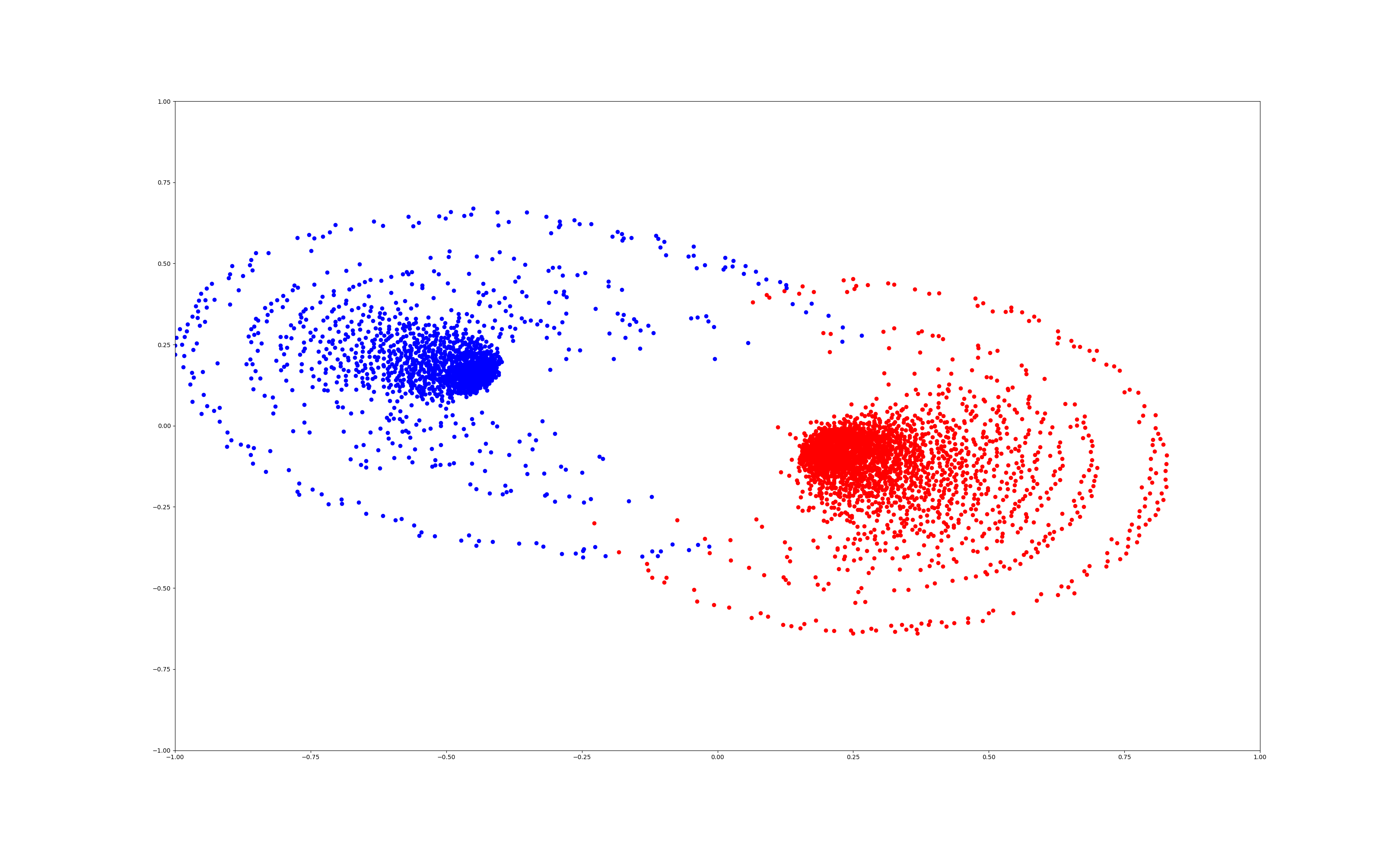}}
         \subfigure[UMAP using R]{\includegraphics[width=0.43\linewidth]{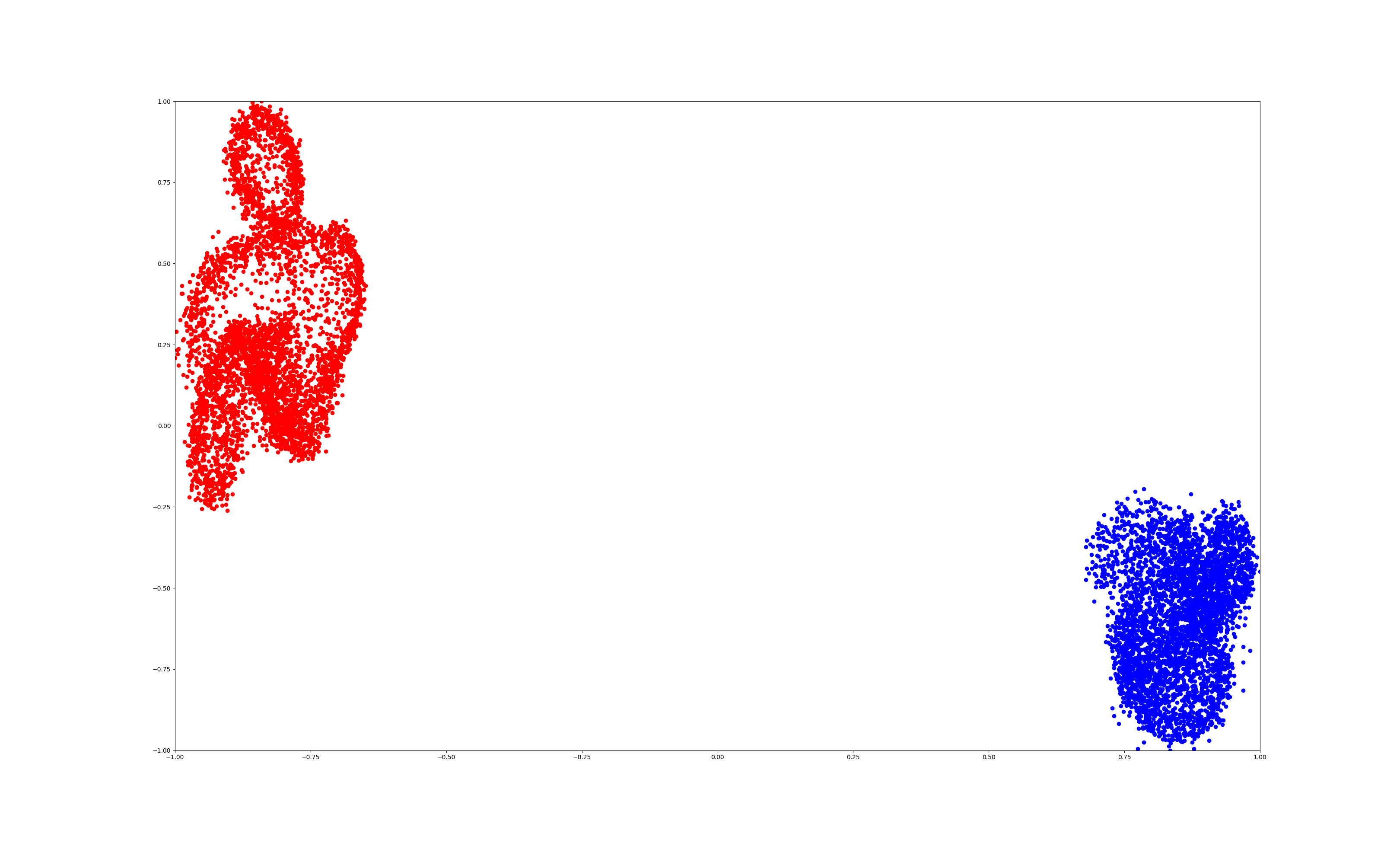}}
         \\
         \subfigure[FD using H]{\includegraphics[width=0.43\linewidth]{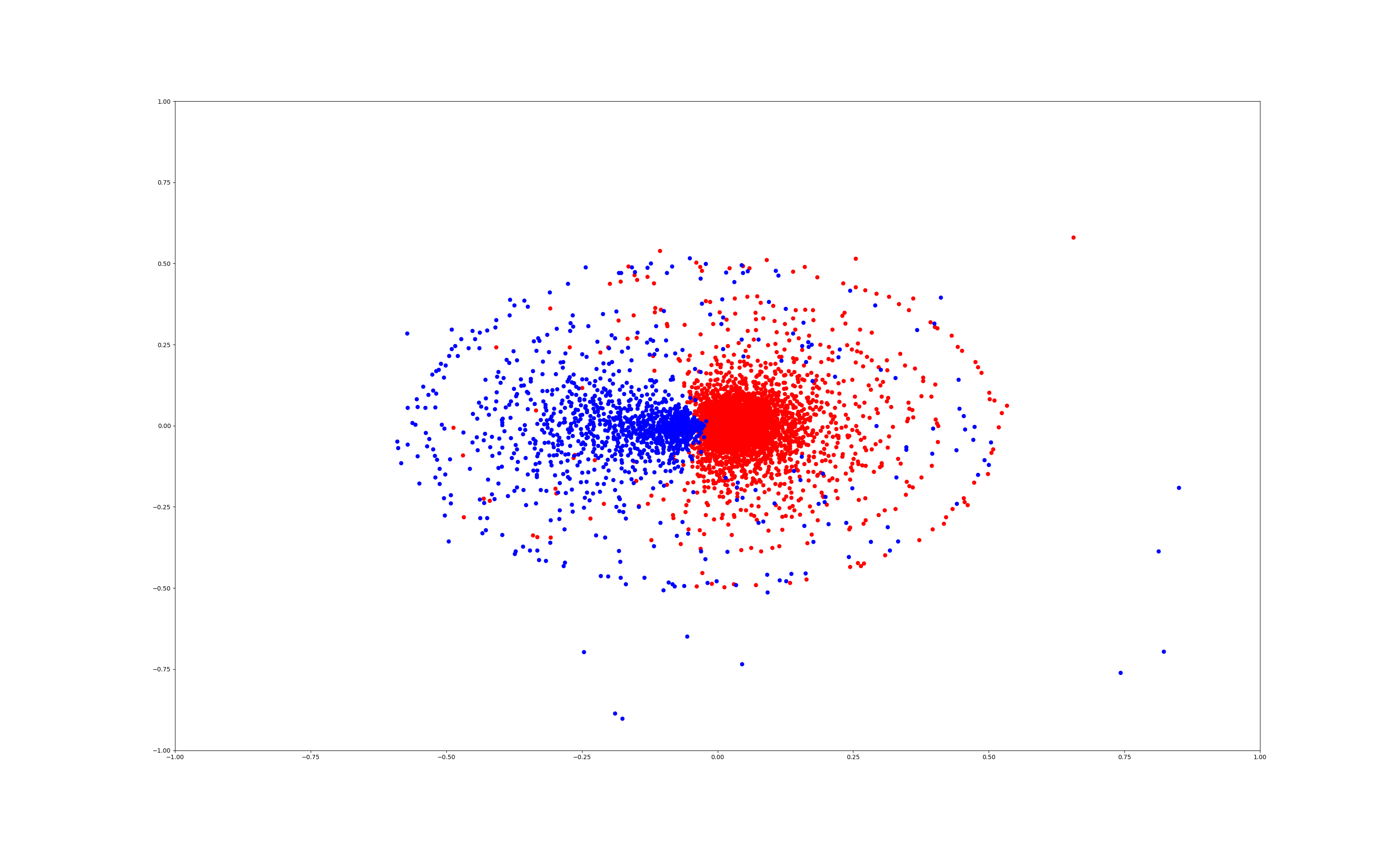}}
         \subfigure[UMAP using H]{\includegraphics[width=0.43\linewidth]{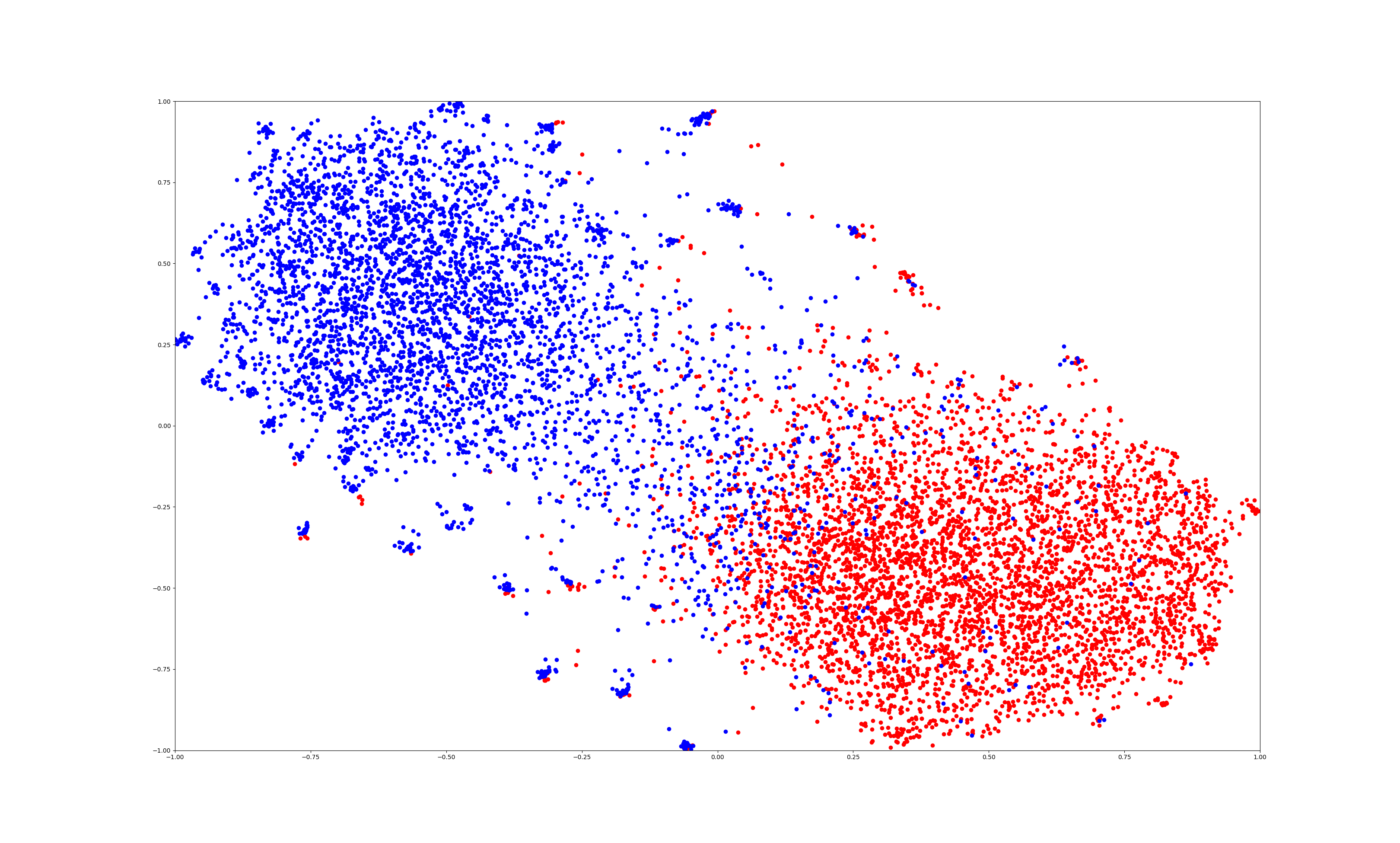}}
    \caption{Comparing FD and UMAP when using Retweets (R) and Hashtags (H). SUPP and OPP users are coded with red and blue respectively.}
    \label{fig:comparingFDandUMAP}
\end{figure}

Figures \ref{fig:resRWC} and \ref{fig:resEC} show the values of RWC and EC using different parameters respectively. As the graphs show, unlike EC, RWC values were fairly stable across different samples sizes and the standard deviation across multiple samples decreased as sample sizes increased.  For RWC and EC and regardless of parameters, users showed higher polarization when we used retweeted accounts to compute similarity compared to when we used hashtags.  This is consistent with prior research which showed that retweets were more indicative of stance than hashtags \cite{magdy2016isisisnotislam}. Using labels from label propagation only led to higher values for both RWC and EC.  This could be attributed to the tendency of label propagation to identify users with more pronounced views.  As for EC, UMAP generally led to higher polarization scores compared to FD.  This is an artifact of the algorithm as it attempts to push dissimilar nodes further apart. Figure \ref{fig:comparingFDandUMAP} illustrates this difference in projecting an identical sets of 5,000 users using FD and UMAP. The figure also shows how users are more separable when computing their similarity using retweets as opposed to hashtags. Further, different samples often led to large standard deviation values for EC.  This indicates that using a single sample to compute EC might not be sufficient.  

To compare RWC without modification (using similarities above a threshold to constitute a link between users) to our modified version, Table \ref{tab:resCompareRWC} compares the unmodified version of RWC at different thresholds with our modified version using a sample of 5,000 users and all user labels for both Retweets and Hashtags.  As the results in the table show, minor modification to the threshold can dramatically change the value of RWC, which is undesirable. 
Given all our analysis, RWC seems to be a more consistent than EC, and when using EC, it is important to compute an average score across multiple user samples.  

\begin{table}[ht!]
    \centering
    \caption{Comparing original RWC to modified RWC}
    \begin{tabular}{c|C{1.5cm}|C{1.5cm}}
Threshold	&	R	&	H	\\ \hline
0.001	&	0.400	&	0.583	\\
0.002	&	0.401	&	0.736	\\
0.003	&	0.687	&	0.896	\\
0.004	&	0.994	&	0.986	\\ \hline
\textbf{modified RWC} & 0.967 & 0.878 \\ 
    \end{tabular}
    \label{tab:resCompareRWC}
\end{table}

Lastly, as Figures \ref{fig:resRWC} and \ref{fig:resEC} show for our BK dataset, users exhibit strong polarization particularly as characterized by the accounts that they retweet (RWC $>$ 0.96 and EC (with UMAP) $>$ 0.62 when using all labeled users). Users' polarization is less pronounced when characterizing them using the hashtag they use (RWC $>$ 0.86 and EC with UMAP $>$ 0.6 when using all labeled users).     

\section{Comparing SUPP and OPP Groups}
After quantifying polarization, we analyzed the data to ascertain the effect of polarization in terms of the differences in interests and focus between both groups as expressed using three elements, namely the hashtags that they use, the accounts they retweet, and the media sources that they cite (share content from). Doing so can provide valuable insights into both groups \cite{darwish2017predicting,darwish2017trump}.  For all three elements, we bucketed them into five bins reflecting how strongly they are associated with the SUPP and OPP groups.  These bins are: strong SUPP, SUPP, Neutral, OPP, and strong OPP.  To perform the bucketing, we used the so-called valence score \cite{conover2011political}, which is computed for an element $e$ as follows:
\vspace{-2pt}
\begin{equation}
    V(e) = 2 \frac{
    \frac{tf_{SUPP}}{total_{SUPP}}}
    {\frac{tf_{SUPP}}{total_{SUPP}} + \frac{tf_{OPP}}{total_{OPP}}} -1
    \label{eq:valence}
\end{equation}
\vspace{-2pt}
where $tf$ is the \textit{frequency} of the element in either the SUPP or OPP tweets and $total$ is the sum of all $tf$s for either the SUPP or OPP tweets. We accounted for all elements that appeared in at least 100 tweets.  Since the value of valence varies between -1 (strong OPP) to +1 (strong SUPP), we divided the range into 5 equal bins: strong OPP [-1.0 -- -0.6), OPP [-0.6 -- -0.2), Neutral [-0.2 -- 0.2), SUPP [0.2 -- 0.6), and strong SUPP [0.6 -- 1.0].

\begin{figure*}[ht]
    \centering
    \subfigure[Hashtags]{\includegraphics[width=0.32\linewidth]{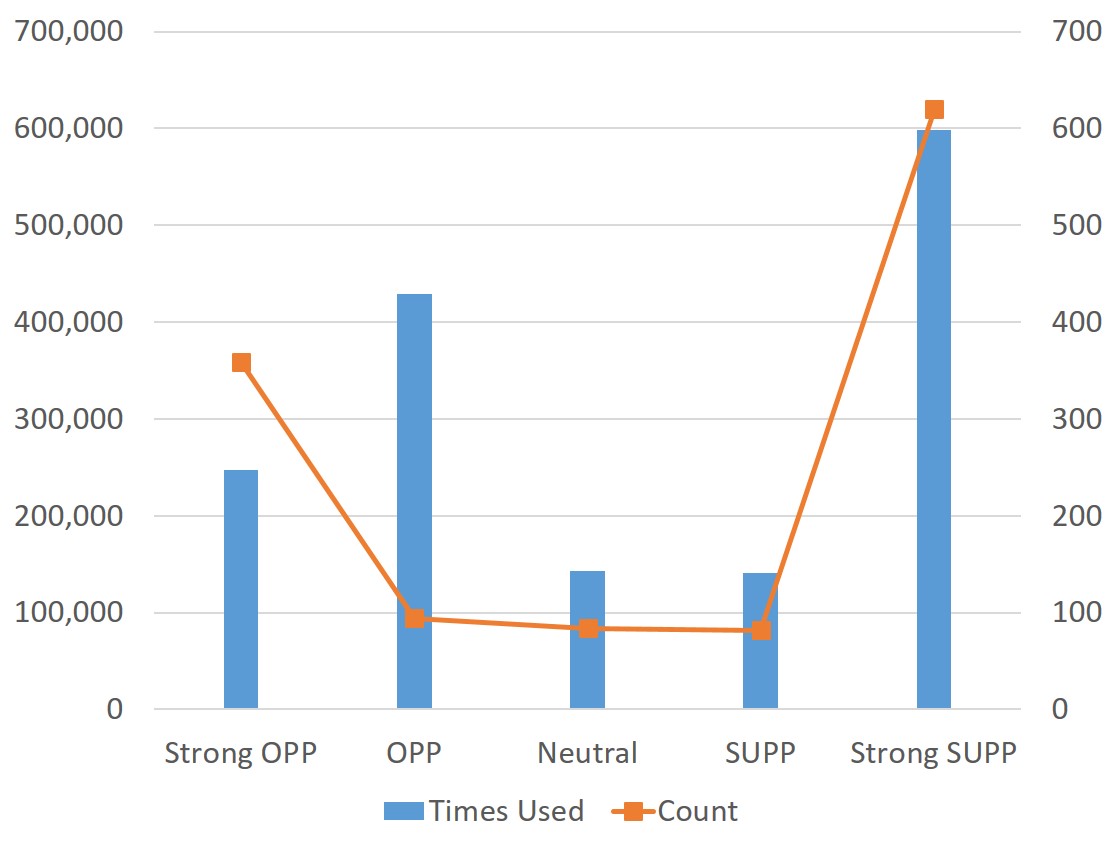}}
    \subfigure[Retweeted Accounts]{\includegraphics[width=0.32\linewidth]{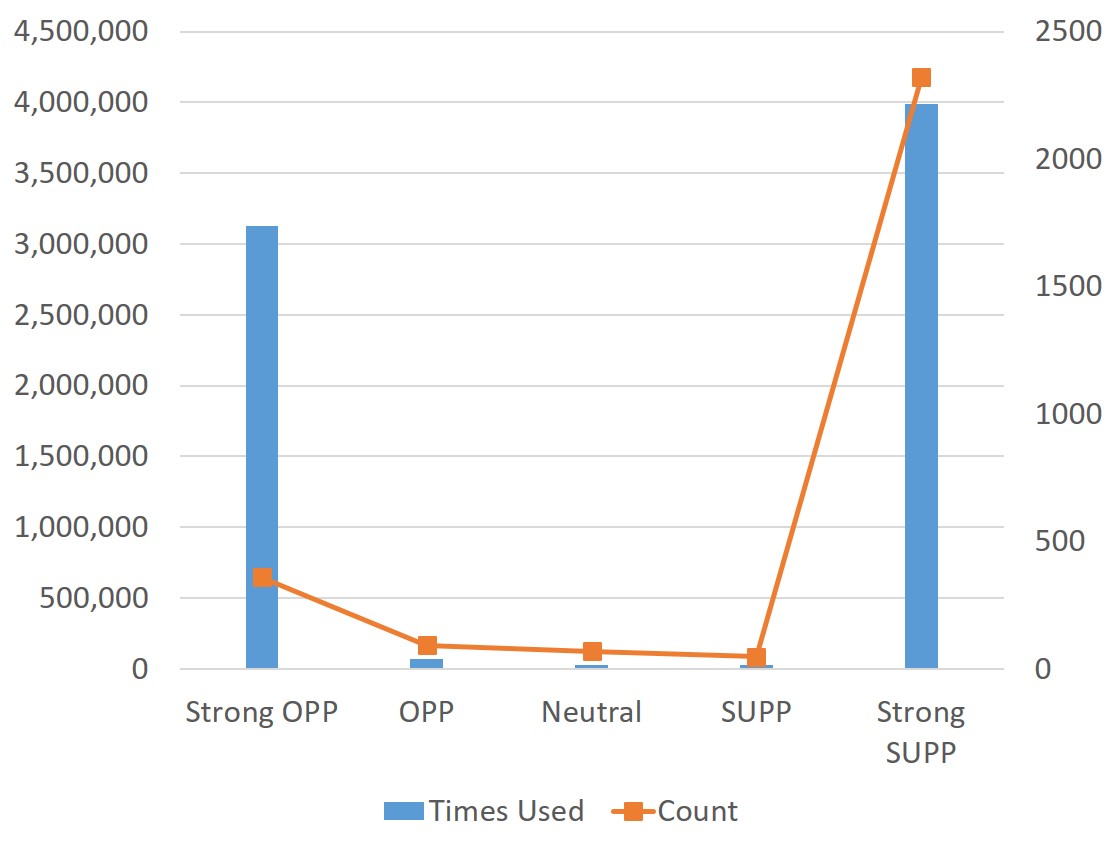}}
    \subfigure[Cited Media]{\includegraphics[width=0.32\linewidth]{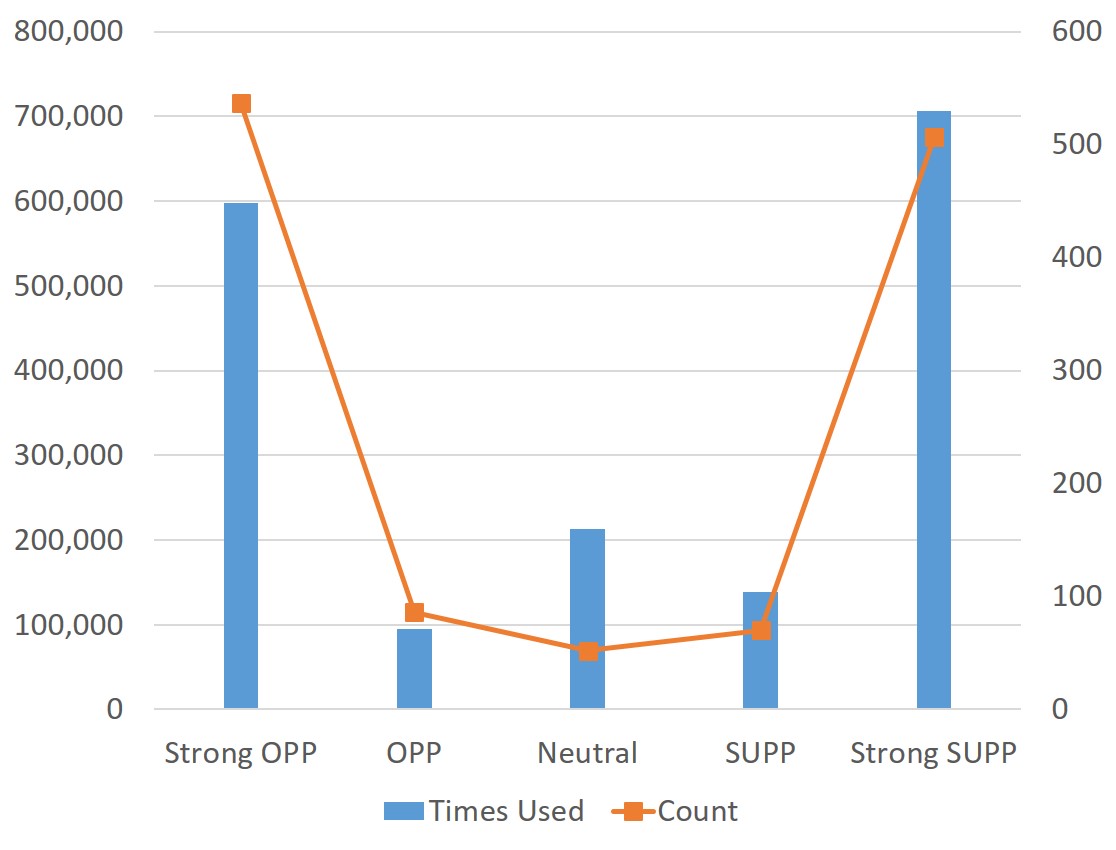}}
    \caption{Count of elements and the number of times they are used for different valence bins}
    \label{fig:URLValence}
\end{figure*}

Figures \ref{fig:URLValence} (a), (b) and (c) respectively provide the number of different hashtags, retweeted accounts, and cited websites that appear for all five bins along with the number of tweets in which they are used.  As the figures show, there is strong polarization between both camps.  Polarization is most evident in the accounts that they retweet and the websites that they share content from, where ``strong SUPP'' and ``strong OPP'' groups dominate in terms of the number of elements and their frequency. This is consistent with the higher values we computed earlier for RWC and EC when using hashtags compared to retweets. If polarization was low, more neutral sources may have been cited more.  Tables \ref{table:topURL} shows the 10 most commonly used hashtags, retweeted accounts, and most cited websites for each of the valence bands.  Since the ``Strong SUPP'' and ``strong OPP'' groups are most dominant, we focus here on their main characteristics.   

\begin{table*}[ht]
\begin{center}
\begin{scriptsize}
\caption{Top 10 elements for each valence band}
    \begin{tabular}{c|c|c|c|c}
    \multicolumn{5}{c}{Top 10 retweeted hashtags} \\
         Strong SUPP	&	SUPP	&	Neutral	&	OPP	&	Strong OPP	\\ \hline
MAGA	&	SCOTUS	&	KavanaughHearings	&	Kavanaugh	&	DelayTheVote	\\
Winning	&	ChristineBlaseyFord	&	KavanaughVote	&	MeToo	&	StopKavanaugh	\\
ConfirmKavanaugh	&	kavanaughconfirmation	&	Breaking	&	BrettKavanaugh	&	GOP	\\
ConfirmKavanaughNow	&	Ford	&	FBI	&	Trump	&	BelieveSurvivors	\\
walkaway	&	kavanaughconfirmed	&	flake	&	republican	&	SNLPremiere	\\
JusticeKavanaugh	&	SaturdayMorning	&	JeffFlake	&	DrFord	&	SNL	\\
QAnon	&	TedCruz	&	SupremeCourt	&	KavanaghHearing	&	TheResistance	\\
Democrats	&	FridayFeeling	&	Grassley	&	BelieveWomen	&	Resist	\\
TXSen	&	HimToo	&	LindseyGraham	&	Republicans	&	Voteno	\\
Midterms	&	TCOT	&	KavanaughHearing	&	RT	&	SusanCollins	\\ \hline\hline

\multicolumn{5}{c}{Top 10 retweeted accounts}\\
         Strong SUPP	&	SUPP	&	Neutral	&	OPP	&	Strong OPP	\\ \hline
realDonaldTrump	&	PollingAmerica	&	RiegerReport	&	AP	&	krassenstein	\\
mitchellvii	&	cspan	&	lachlan	&	CBSNews	&	kylegriffin1	\\
dbongino	&	JenniferJJacobs	&	Sen\_JoeManchin	&	Reuters	&	KamalaHarris	\\
charliekirk11	&	JerryDunleavy	&	AaronBlake	&	USATODAY	&	SenFeinstein	\\
FoxNews	&	JulianSvendsen	&	WSJ	&	Phil\_Mattingly	&	EdKrassen	\\
RealJack	&	jamiedupree	&	markknoller	&	dangercart	&	thehill	\\
DineshDSouza	&	CNNSotu	&	Bencjacobs	&	WalshFreedom	&	MichaelAvenatti	\\
Thomas1774Paine	&	AlBoeNEWS	&	lawrencehurley	&	4YrsToday	&	SethAbramson	\\
AnnCoulter	&	elainaplott	&	AureUnnie	&	byrdinator	&	funder	\\
foxandfriends	&	AlanDersh	&	choi\_bts2	&	MittRomney	&	Lawrence	\\ \hline\hline

\multicolumn{5}{c}{Top 10 media sources (T stands for twitter.com)}\\
         Strong SUPP	&	SUPP	&	Neutral	&	OPP	&	Strong OPP	\\ \hline
thegatewaypundit.com	&	usatoday.com	&	dr.ford	&	nytimes.com	&	hill.cm	\\
foxnews.com	&	mediaite.com	&	T/michaelavenatti/	&	T/thehill/	&	washingtonpost.com	\\
dailycaller.com	&	T/realdonaldtrump/	&	dailym.ai	&	thehill.com	&	rawstory.com	\\
breitbart.com	&	theweek.com	&	lawandcrime.com	&	politi.co	&	vox.com	\\
thefederalist.com	&	T/lindseygrahamsc/	&	nypost.com	&	abcn.ws	&	huffingtonpost.com	\\
westernjournal.com	&	nyp.st	&	T/donaldjtrumpjr/	&	usat.ly	&	nyti.ms	\\
politico.com	&	T/senfeinstein/	&	T/senjudiciary/	&	axios.com	&	nbcnews.com	\\
ilovemyfreedom.org	&	T/kamalaharris/	&	T/mediaite/	&	politico.com	&	CNN.com	\\
chicksonright.com	&	T/newsweek/	&	c-span.org	&	reut.rs	&	apple.news	\\
hannity.com	&	T/natesilver538/	&	T/gop/	&	po.st	&	dailykos.com	\\
    \end{tabular}
    \label{table:topURL}
    \end{scriptsize}
\end{center}
\end{table*}

For the \textbf{``Strong SUPP''} group, the hashtags can be split into the following topics (in order of importance as determined by frequency):
\begin{itemize}
    \item \textbf{Trump related:} \#MAGA (Make America Great Again), \#Winning.
    \item \textbf{Pro BK confirmation:} \#ConfirmKavanaugh, \#ConfirmKavanaughNow, \#JusticeKavanaugh.
    \item \textbf{Anti-DNC:} \#walkAway (from liberalism), \#Democrats, \#Feinstein
    \item \textbf{Conspiracy theories:} \#QAnon (an alleged Trump administration leaker), \#WWG1WGA (Where We Go One We Go All)
    \item \textbf{Midterm elections:} \#TXSen (Texas republican senator Ted Cruz), \#Mid-terms, \#VoteRed2018 (vote Republican)
    \item \textbf{Conservative media:} \#FoxNews, \#LDTPoll (Lou Dobbs (FoxNews) on Twitter poll)
\end{itemize}
It is interesting to see hashtags expressing support for Trump (\#MAGA and \#Wining) feature more prominently than those that indicate support for BK. 
Retweeted accounts for the same group reflect a similar trend: 
\begin{itemize}
    \item \textbf{Trump related:} \@realDonaldTrump, \@mitchellvii (Bill Mitchell: social media personality who staunchly supports Trump), \@RealJack (Jack Murphy: co-owner of \url{ILoveMyFreedom.org} (pro-Trump website)), \@DineshDSouza (Dinesh D'Souza: commentator and film maker)
    \item \textbf{Conservative media:} \@dbongino (Dan Bongino: author with podcast), \@FoxNews, \@FoxAndFriends (Fox News), \@JackPosobiec (Jack Posobiec: One America News Network), \@IngrahamAngle (Laura Ingraham: Fox News)
    \item \textbf{Conservative/GOP personalities:} \@charliekirk11 (Charlie Kirk: founder of Turning Point USA), \@AnnCoulter (Ann Coulter: author and commentator), \@Thomas1774Paine (Thomas Paine: author), \@paulsperry\_ (Paul Sperry: author and media personality), \@RealCandaceO (Candace Owens: Turning Point USA), \@McAllisterDen (D. C. McAllister: commentator)
\end{itemize}
The list above show that specifically pro-Trump accounts featured even more prominently than conservative accounts. Table 4 
lists cited media for the ``Strong SUPP'' group.  The media were generally right-leaning, with some of them being far-right and most of them having mixed credibility.  

\begin{table}[ht!]
\begin{center}
\caption{Top cited media for SUPP and OPP groups.  Bias and credibility are determined by MediaBiasFactCheck.com.}
\footnotesize
\begin{tabular}{l|c|c||l|c|c}
\multicolumn{3}{c||}{Strong SUPP} &  \multicolumn{3}{c}{Strong OPP} \\
Source & Bias & Cred. & Source & Bias & Cred. \\ \hline
theGatewayPundit.com & far right & low & theHill.com & left-center & high\\
FoxNews.com & right & mixed & WashingtonPost.com & left-center & high\\
DailyCaller.com & right & mixed & RawStory.com & left & mixed \\
breitbart.com & far right & low & Vox.com & left & high\\
theFederalist.com & right & high & HuffingtonPost.com & left & high\\
WesternJournal.com & right & mixed & NYTimes & left-center & high\\
Politico.com & left-center & high & NBCNews.com  & left-center & high\\
ILoveMyFreedom.org & far right & low & CNN.com & left & mixed \\
ChicksOnRight.com & -- & -- & apple.news & -- & --\\
Hannity.com (FoxNews) & -- & -- & dailykos.com & left & mixed\\
\end{tabular}
\end{center}
\label{tab:mostCitedMedia}
\end{table}

For the \textbf{``strong OPP''}, the top hashtags can be topically grouped as:
\begin{itemize}
    \item \textbf{Anti-BK:} \#DelayTheVote, \#StopKavanaugh, \#KavaNo (BK no), \#voteNo.
    \item \textbf{Republican Party related:} \#GOP, \#SusanCollins (GOP senator voting for BK).
    \item \textbf{Sexual assault related:} \#BelieveSurvivors, \#JulieSwetnick (BK accuser).
    \item \textbf{Media related:} \#SNLPremiere (satirical show), \#SNL, \#SmartNews (anti-Trump/GOP news)
    \item \textbf{Anti Trump:} \#TheResistance, \#Resist
    \item \textbf{Midterms:} \#vote, \#voteBlue (vote democratic)
\end{itemize}
As the list shows, the most prominent hashtags were related to opposition to the confirmation of BK.  Opposition to the Republican Party (\#GOP) and Trump (\#TheResistance) may indicate polarization.

As for their retweeted accounts, media related accounts dominated the list.  The remaining accounts belonged to prominent Democratic Party officials and anti-Trump accounts.  The details are as follows (in order of importance):



\begin{itemize}
    \item \textbf{Media related:} \@krassenstein (Brian Krassenstein: \url{HillReporter.com}), \@kylegriffin1 (Kyle Griffin: MSNBC producer), \@theHill, \@EdKrassen (Ed Krassenstein: \url{HillReporter.com}), \@funder (Scott Dworkin: Dworkin Report and Democratic Coallition), \@Lawrence (Lawrence O'Donnell: MSNBC), \@MSNBC, \@JoyceWhiteVance (Joyce Alene: professor and MSNBC contributor), \@Amy\_Siskind (Amy Siskind: The Weekly List)
    \item \textbf{DNC:} \@KamalaHarris (Senator), \@SenFeinstein (Senator Dianne Feinstein), \@TedLieu (Representative)
    \item \textbf{Anti Kavanaugh:} \@MichaelAvenatti (lawyer of BK accuser)
    \item \textbf{Anti Trump:} \@SethAbramson (author of ``Proof of Collusion''), \@tribelaw (Laurence Tribe: Harvard Professor and author of ``To End a Presidency'')
\end{itemize}

Concerning the cited media shown in Table 4, 
they were mostly left or left-of-center leaning sources.  The credibility of the sources were generally higher than those for the ``strong SUPP'' group.  

\section{Conclusion}

In this paper, we characterized the political polarization on Twitter concerning the nomination of judge Brett Kavanaugh to the US Supreme Court.  We used the automatically tagged set of more than 128 thousand Twitter users to ascertain the robustness of two different measures of polarization quantification.  We proposed changes to both measures to make them more efficient and more effective.  We showed that those who support and oppose the confirmation of Kavanaugh were generally using divergent hashtags and were following different Twitter accounts and media sources. For future work, we plan to look at different topics with varying levels of polarization, as the Kavanaugh nomination was a strongly polarizing topic.   




\bibliographystyle{splncs04} 
\bibliography{references}

\begin{thebibliography}{10}
\providecommand{\url}[1]{\texttt{#1}}
\providecommand{\urlprefix}{URL }
\providecommand{\doi}[1]{https://doi.org/#1}

\bibitem{borge2015content}
Borge-Holthoefer, J., Magdy, W., Darwish, K., Weber, I.: Content and network
  dynamics behind egyptian political polarization on twitter. In: Proceedings
  of the 18th ACM Conference on Computer Supported Cooperative Work \& Social
  Computing. pp. 700--711. ACM (2015)

\bibitem{conover2011political}
Conover, M., Ratkiewicz, J., Francisco, M.R., Gon{\c{c}}alves, B., Menczer, F.,
  Flammini, A.: Political polarization on twitter. Icwsm  \textbf{133},  89--96
  (2011)

\bibitem{darwish2017predicting}
Darwish, K., Magdy, W., Rahimi, A., Baldwin, T., Abokhodair, N.: Predicting
  online islamophopic behavior after\# parisattacks. The Journal of Web Science
   \textbf{3}(1) (2017)

\bibitem{darwish2017trump}
Darwish, K., Magdy, W., Zanouda, T.: Trump vs. hillary: What went viral during
  the 2016 us presidential election. In: International Conference on Social
  Informatics. pp. 143--161. Springer (2017)

\bibitem{darwish2019unsupervisedStance}
Darwish, K., Stefanov, P., Aupetit, M.J., Nakov, P.: Unsupervised user stance
  detection on twitter. arXiv preprint arXiv:1904.02000  (2019)

\bibitem{garimella2018quantifying}
Garimella, K., Morales, G.D.F., Gionis, A., Mathioudakis, M.: Quantifying
  controversy on social media. ACM Transactions on Social Computing
  \textbf{1}(1), ~3 (2018)

\bibitem{guerra2013measure}
Guerra, P.C., Meira~Jr, W., Cardie, C., Kleinberg, R.: A measure of
  polarization on social media networks based on community boundaries. In:
  Seventh International AAAI Conference on Weblogs and Social Media (2013)

\bibitem{joulin2016bag}
Joulin, A., Grave, E., Bojanowski, P., Mikolov, T.: Bag of tricks for efficient
  text classification. arXiv preprint arXiv:1607.01759  (2016)

\bibitem{kutlu2018devam}
Kutlu, M., Darwish, K., Elsayed, T.: Devam vs. tamam: 2018 turkish elections.
  arXiv preprint arXiv:1807.06655  (2018)

\bibitem{magdy2016isisisnotislam}
Magdy, W., Darwish, K., Abokhodair, N., Rahimi, A., Baldwin, T.: \#
  isisisnotislam or\# deportallmuslims?: Predicting unspoken views. In:
  Proceedings of the 8th ACM Conference on Web Science. pp. 95--106. ACM (2016)

\bibitem{magdy2016failedrevolutions}
Magdy, W., Darwish, K., Weber, I.: \# failedrevolutions: Using twitter to study
  the antecedents of isis support. First Monday  \textbf{21}(2) (2016)

\bibitem{mohtarami2018automatic}
Mohtarami, M., Baly, R., Glass, J., Nakov, P., M{\`a}rquez, L., Moschitti, A.:
  Automatic stance detection using end-to-end memory networks. In: Proceedings
  of the 2018 Conference of the North American Chapter of the Association for
  Computational Linguistics: Human Language Technologies, Volume 1 (Long
  Papers). vol.~1, pp. 767--776 (2018)

\bibitem{morales2015measuring}
Morales, A., Borondo, J., Losada, J.C., Benito, R.M.: Measuring political
  polarization: Twitter shows the two sides of venezuela. Chaos: An
  Interdisciplinary Journal of Nonlinear Science  \textbf{25}(3),  033114
  (2015)

\bibitem{weber2013secular}
Weber, I., Garimella, V.R.K., Batayneh, A.: Secular vs. islamist polarization
  in egypt on twitter. In: Proceedings of the 2013 IEEE/ACM International
  Conference on Advances in Social Networks Analysis and Mining. pp. 290--297.
  ACM (2013)

\end{thebibliography}

\end{document}